\documentclass[review]{elsarticle}

\usepackage{hyperref}
\usepackage{amsmath,amssymb}

\usepackage{color}
\usepackage{changepage}
\usepackage{diagbox}
\usepackage{makecell}
\usepackage{amsthm}
\newtheorem{definition}{Definition}

\def\so {\textbf{SocNet}} 
\def\eso {\textbf{ESocNet}} 
\def\bso {\textbf{BSocNet}} 
\def\se {\textbf{SexNet}} 
\def\sts {$\textbf{Soc2Sex}~$} 

\def\defeqq{:=}
\journal{Journal of \LaTeX\ Templates}

\usepackage{fixltx2e}
\usepackage{algpseudocode}
\usepackage{mathtools}  
\usepackage{amsmath}
\usepackage{amssymb}
\usepackage{tabulary}
\usepackage{rotating}

\usepackage{colortbl} 
\usepackage{xcolor} 
\usepackage{xfrac}
\usepackage{graphicx}
\usepackage{subcaption}
\usepackage{array}
\usepackage{xcolor} 
\usepackage{dcolumn}
\usepackage{tabu}
\usepackage{array}
\usepackage{multirow}
\usepackage{tikz}
\usetikzlibrary{arrows.meta,positioning}
\usetikzlibrary{calc}
\usetikzlibrary{snakes}

\usepackage{dot2texi}
\usepackage[ruled,vlined]{algorithm2e}
\usepackage{booktabs}

\begin{document}

\begin{frontmatter}

\title{Generating a Heterosexual Bipartite Network Embedded in Social Network}

\author[mymainaddress1]{Asma Azizi\corref{mycorrespondingauthor}}
\cortext[mycorrespondingauthor]{Corresponding author}
\ead{aazizibo@uci.edu }
\address[mymainaddress1]{Department of Mathematics; University of California, Irvine, CA 92612}

\author[mysecondaryaddress2]{Zhuolin Qu}
\address[mysecondaryaddress2]{ Department of Mathematics
The University of Texas at San Antonio, San Antonio, Texas 78249}

\author[mysecondaryaddress3]{Bryan Lewis}
\address[mysecondaryaddress3]{ Network Dynamics and Simulation Sciences Laboratory, Virginia Bioinformatics Institute at Virginia Tech,
Blacksburg, VA 24061, U.S.A}

\author[mysecondaryaddress4]{James Mac Hyman}
\address[mysecondaryaddress4]{Department of Mathematics, Tulane University, 
New Orleans, LA 70118, U.S.A}


\begin{abstract}
~\\
We describe how to generate a  heterosexual network with a prescribed joint-degree distribution that is embedded in a prescribed large-scale social contact network.   
The structure of a sexual network plays an important role in how sexually transmitted infections (STIs) spread. Generating an ensemble of networks that mimics the real-world is crucial to evaluating robust mitigation strategies for controling STIs.
Most of the current algorithms to generate sexual networks only use sexual activity data, such as the number of partners per month, to generate the sexual network.  
Real-world sexual networks also depend on biased mixing based on age, location, and social and work activities. We describe an approach to use a broad range of social activity data to generate possible heterosexual networks.
We start with a  large-scale simulation of thousands of people in a city as they go through their daily activities, including work, school, shopping, and  activities at home. 
We extract a social network from these activities where the nodes are the people and the edges indicate a social interaction, such as working in the same location.  
This social network captures the correlations between people of different ages, living in different locations, their economic status, and other demographic factors.  
We  use the social contact network to define a  bipartite heterosexual network that is embedded within an extended social network.
The resulting sexual network captures the biased mixing inherent in the social network, and models based on this pairing of networks can be used to investigate novel intervention strategies based on the social contacts of infected people. 
We illustrate the approach in a model for the spread of Chlamydia in the heterosexual network representing the young sexually active community in New Orleans.

\end{abstract}


\end{frontmatter}

\section{Introduction}\label{Introduction}
The  structure of heterosexual networks plays an important role in the spread of sexually transmitted infections (STIs). These networks are captured in computer simulations by a bipartite graph where the nodes represent the people and the edges are sexual partnerships between nodes of different sexes.  Determining what is predictable in STI models requires an algorithm to generate an ensemble of random graphs that resembles real-world sexual activities, including the distribution for the number of sexual partners people have (their degree distribution)  and the number of partners their partners have (the joint-degree, or degree-degree, distribution). 
The existing algorithms that generate bipartite random graphs preserving degree and joint-degree distributions of the nodes are strictly based  on the number of partners people have, and not other demographic factors, such as age or location \cite{newman2002structure, hakimi1962realizability,boroojeni2017generating}.

The degree and joint-degree distributions are just two of many properties for a heterosexual network that can affect its  structure and the validity of an epidemic model. The heterosexual network is also correlated to an underlying \textit{social contact network} of acquaintances connected by interpersonal relationships.  A person's sexual activity depends on age, race, sociodemographic, and socioeconomic features of the environment that can be captured by a social contact network   \cite{amirkhanian2014social,adimora2005social,ruan2011sexual,juher2017network, morris1995social}. 
In other words, using the extended social network of a person as a source of sexual partner selection  when generating a heterosexual network, enables the network to capture the bias in heterogeneous mixing based on age, race, economic status, and geographic location  \cite{mcpherson2001birds}. 

Although it is widely accepted that social contact (non-sexual partners) and heterosexual (sexual partners) networks are related,  there are few studies on how a population's social contact network impacts the spread of heterosexual STIs. 
The social network  can affect the structure of the heterosexual network by providing the pool of sexual partners and  can transmit information and cultural norms regarding STI test and safer sex, but  there is no mechanistic approach that addresses and uses this capability of social networks. We will describe a new approach that fills this gap by applying social contact networks  to generate the heterosexual network while preserving the joint-degree distribution of data. 

Our new network generation approach uses the underlying extended social network of a population to extend these previous algorithms for generating bipartite heterosexual networks with prescribed joint-degree distribution \cite{boroojeni2017generating}.
Many sexual partnerships are formed from within a person's social circle, defined by the people they have regular social contact with, and the contacts of their contacts (their extended social network).  
These social circles have been modeled through large-scale simulations of thousands of people in a city as they go through their daily activities. 
We start with a network that mimics the social activity of the population \cite{eubank2010detail}, as generated by a complex social network simulation.
 We use this simulated data to create an extended social network and then identify a bipartite network of men and women to  define our heterosexual network. 
We then create a virtual heterosexual network as a subgraph of this bipartite social network that captures a prescribed  joint-degree  distribution.

As a case study, we construct a heterosexual network that is embedded in the social contact network of New Orleans population and  mimics the sexual behavior obtained from a sexual behavior survey of the young adult African American population in New Orleans \cite{kissinger2014check,green2014influence}.

\section{Materials and method}
\label{method}

People often find their sexual partners within their extended social network, the individuals they come in contact with each day at work, school, or other social activities. 
There are sophisticated simulations of these social networks that can be used to produce a sexual network, which  is more realistic than basing partnerships on just the sexual activity of different individuals. 

The social contact network  is a graph where the nodes are synthetic people, labeled by their demographics (sex, age, income, location, etc.),
and the edges between the nodes represent contacts determined in which each synthetic person is deemed to have made contact with a subset of other synthetic people through some \textit{Activity} types.  Each edge of the network is labeled with one of these activity locations and is weighted by  the time spent on these contacts per day. For example edge (\textbf{i},\textbf{j}) labeled by the activity $A=Work$ and weighted by  $T^W_{i,j}$   means two persons \textbf{i} and \textbf{j} have a contact for  $T^W_{i,j}$ fraction of their total time  spent at work.  
 We base our algorithm on a social contact network, called \so,  generated by Eubank et al.  \cite{eubank2010detail} with activity at different locations (e.g. home, work, school, shopping, or other activity).

We introduce an algorithm that embeds a heterosexual  network within a social network and also meets the joint-degree distribution of the sexually active population. 
The heterosexual network preserves the bipartite joint degree ($BJD$) distribution matrix that represents the correlations between then number of partners a person has and the number of partners their partners have \cite{boroojeni2017generating}. 
The   algorithm has three stages:
\begin{enumerate}[(i)]
\item  \underline{Generate an extended social contact network, \eso:}  The original social contact network is a simple graph, whose nodes are synthetic people, and neighboring nodes are their social contacts during a typical day.  We assume that most sexual partnerships come from a person's social contacts, or the social contacts of their social contacts, e.g. the neighbors of the neighbors of a node.  We extend the social contact network  to create a new network, the extended social network,  \eso, where some of the neighbors (social contacts) of an individual's neighbors in this network are added to his/her social contacts. 

\item \underline{Generate a reduced social bipartite network, \bso:} The \eso~ includes all the individuals in the region being modeled. Our sexual network is based on individuals within a prescribed age range.  In this step, we remove all nodes where the associated individuals are outside this age range.  
The extended social network is a simple graph where nodes have some neighbors that are the same sex.  We identify the embedded bipartite subgraph of this network by removing all edges between individuals of the same sex.  The resulting bipartite graph is a social network where male nodes are only connected to female nodes and vice versa.  Finally, we assume siblings are not sexual partners and therefore, we remove all edges between individuals living in the same household, which is the edge labeled activity H for home. We call this reduced social bipartite network as  \bso.

\item \underline{Generate an embedded heterosexual bipartite network, \se:}  We then use the \bso~ to define a heterosexual network of sexual partnerships, the \se,  with a prescribed $BJD$ based on survey data \cite{boroojeni2017generating}.  
That is, we preserve the correlations between the number of partners a person has and the distribution for the number of partners their partners have. 
We assume that most of a person's sexual partners are neighbors in the \bso~ and a few of the partners are randomly selected from elsewhere in the population where they might have met through social media or at any other event. 
\end{enumerate}

\subsection{Generate an extended social contact network (\eso)}
In the first stage of our algorithm, we create an extended social contact network, \eso,  so that an individual's social contacts include some of the contacts of their contacts.  
That is,  \eso~ will add potential sexual partners by including some of the social contacts of an individual's social contacts. 

Consider two people (nodes) \textbf{i} and \textbf{j} who are not currently connected, but have $k_A(i,j)>0$ common social contacts within activity $A$. 
We define $p^A_{ij}$ as the probability that they will meet through a single contact. Therefore, the probability that they  will meet and be connected in the \eso~  after $k_A(i,j)$ contacts is  $1-(1-p^A_{ij})^{k_A(i,j)}$.  

\begin{figure}[htp]
\centering
  \begin{tikzpicture}[
      mycircle/.style={
         circle,
         draw=black,
         fill=gray,
         fill opacity = 0.3,
         text opacity=1,
         inner sep=0pt,
         minimum size=20pt,
         font=\small},
      myarrow/.style={-},
      node distance=1.5cm and 1.9cm
      ]
      \node[mycircle] (i) {$\textbf{i}$};
      \node[mycircle, right=of i] (k1) {\textbf{k}$_1$};
      \node[mycircle,above =of k1] (k0) {\textbf{k}$_0$};
       \node[mycircle,below =of k1] (k2) {\textbf{k}$_2$};
       \node[mycircle,right =of k1] (j) {\textbf{j}};
      
     \draw [myarrow] (i) -- node[sloped,above] {$A_{ik_0}$,~$T^A_{ik_0}$} (k0);
     \draw [myarrow] (i) -- node[sloped,above] {$A_{ik_1}$,~$T^A_{ik_1}$} (k1);
     \draw [myarrow] (i) -- node[sloped,above] {$A_{ik_2}$,~$T^A_{ik_2}$} (k2);
     \draw [myarrow] (j) -- node[sloped,above] {$A_{jk_0}$,~$T^A_{jk_0}$} (k0);
     \draw [myarrow] (j) -- node[sloped,above] {$A_{jk_1}$,~$T^A_{jk_1}$} (k1);
     \draw [myarrow] (j) -- node[sloped,above] {$A_{jk_2}$,~$T^A_{jk_2}$} (k2);
    \end{tikzpicture}
    \vspace*{8pt}
    \caption{Suppose the persons \textbf{i} and \textbf{j}  are not currently  social contacts in \so  but have three different common social contacts \textbf{k}$_0$,\textbf{k}$_1$, and \textbf{k}$_2$ through different activities.
They might be connected in the extended social network, \eso,  when at least one of their common social contacts meet them within the same activity location.
Suppose $A_{ik_0}=A_{jk_0}=A_{ik_1}=A_{jk_1}=A\neq A_{ik_2}\neq A_{ik_2}$, that is, \textbf{k}$_0$ meets \textbf{i} and  \textbf{j} at the same location, similarly \textbf{k}$_1$ meets \textbf{i} and  \textbf{j} at the same location to \textbf{k}$_0$'s, however, \textbf{k}$_2$ meets them in different places. To compute  $p^A_i$ and $p^A_j$ we only count the social contacts who meet them at the same location A, therefore, $p^A_i=(T^A_{ik_0}+T^A_{ik_1})/{2}$, and $p^A_j=(T^A_{jk_0}+T^A_{jk_1})/{2}$. 
Finally, the probability that \textbf{i} and \textbf{j} make an edge- are connected in the extended social network- is $1-(1-p^A_ip^A_j)^{2}.$
}
    \label{frined_of_friend}
\end{figure}
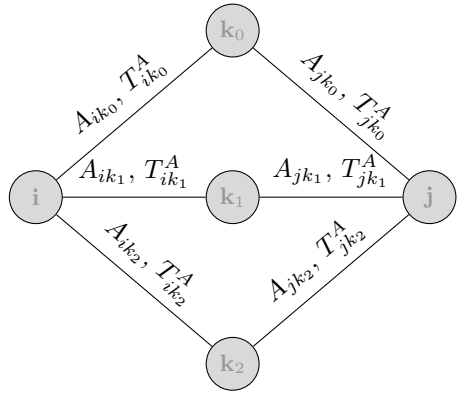

The probability $p^A_{ij}$ is a function of the time that \textbf{i} and \textbf{j} spend in an activity $A$ in \so.
From the data in \so, we can define $\tau^A_k$ as the average fraction of time person \textbf{k} spends with each  social contact, when engaged in  activity $A$:   
\begin{eqnarray}
\tau^A_k=\frac{\sum_{l \in N_A(k) } T^A_{k,l}}{|N_A(k)|}~,
\end{eqnarray}
where $T^A_{k,l}$ is the fraction of time two contacts \textbf{k} and \textbf{l} spend together in activity $A$, and $N_A(k)$ is set of all  social contacts for person \textbf{k} through an activity location $A$.
We then define $p^A_{ij}=\tau^A_i \tau^A_j$. 
Figure (\ref{frined_of_friend}) describes  an schematic of this algorithm for a simple network. 
 
\subsection{Generate a  reduced social bipartite network, \bso}
In our heterosexual network,  we only consider the sexually active population within a prescribed age range $\alpha=[\alpha_1,\alpha_2]$.  That is, we trim the \eso~ by removing all people with ages outside this range to  not including any edges (sexual contacts) with people outside this range.  
We then remove all edges between people of the same sex to create a bipartite heterosexual social  network.  
Finally, to avoid including siblings as potential sexual partners, we remove all edges between two individuals living in the same household by removing all edges labeled activity H as home, to define the \bso.
\subsection{Generate an embedded heterosexual network (\se)}
The \sts algorithm uses \bso~  to generates a heterosexual network,  \se, that mimics the heterogeneous mixing of the real population. 
We assume that we have an estimate for the distribution for the number of partners of men and women (the degree distributions for their associated nodes) and the joint-degree distribution for the number of partners that their partners have \cite{boroojeni2017generating}.

An edge, $\textbf{ij}$,  between two persons \textbf{i} and \textbf{j}  in \se~ represents a sexual partnership. 
The \textit{degree}  of a person \textbf{i}, is defined by the number of his/her sexual partners.
The \textit{degree distribution} $\{d_k\}$ defines the number of people with  degree $k$.
The \textit {joint-degree distribution}  $(k,j)$ is the number of partnerships between a man with degree $j$ and a woman with degree $k$.
This distribution  can be represented by the \textit{Bipartite Joint Degree} or $BJD$ matrix:

\[ 
    BJD_{\se}=\begin{pmatrix} 
 e_{11} & e_{12} & e_{13} & \dotsm  & e_{1m}\\
e_{21} & e_{22} & e_{23} & \dotsm  & e_{2m}\\
\multicolumn{5}{c}{$\vdots$}     \\
e_{w1} & e_{w2} & e_{w3} & \dots  & e_{wm}
    \end{pmatrix}, 
\]
where, $w$ is the maximum degree in women nodes, and $m$ is the maximum degree in men nodes, each element $e_{ij}$ is the number of edges between women with $i$ partners and men with $j$ partners. 

The degree distribution of the number of women nodes, $d^w_k$, and men nodes, $d^m_k$, with  $k$ partners can be obtained from $BJD_{\se}$:
\begin{eqnarray}
d^w_k=\frac{\sum_{j=1}^m e_{kj}}{k}~,\text{~~and~~~}~~~d^m_k=\frac{\sum_{i=1}^w e_{ik}}{k}~.
\end{eqnarray}

Though the heterosexual network  \se~  is a subgraph of  \bso, we  also consider some sexual partners that are within a person's extended social circle.  That is, for the general case, \se~ is partially embedded in \bso.

The \sts algorithm first generates an initial heterosexual network that closely agrees with the desired $BJD$ matrix and is partially embedded in  \bso.  Usually, this network satisfies the desired $BJD$, but can fail when the average degree (number of partners people have) becomes large.  When this happens, a second fix-up algorithm, based on rewiring  the network, is used to repair any discrepancies so the final \se~ has the desired $BJD$ matrix.
\subsubsection{Generating the bipartite network }
The \sts algorithm starts with the \textbf{SocNet}, the $BJD$ matrix corresponding to \textbf{SexNet}, and the fraction $p\in[0,1]$ of partners that are chosen randomly from the extended social contacts in the \bso. 
The remaining fraction, $(1-p)$, of a person's sexual partners are randomly selected from elsewhere in the population. 
These partnerships might have formed by meeting through social media or a social event not captured by the original \so. 
The \sts algorithm then generates a heterosexual network that is a partial subgraph of \bso~ and has a joint-degree distribution given by the $BJD$ matrix.
Note that $p$ is approximately the percentage of \se~ that is a subgraph of \bso. 

The algorithm starts with an empty set of  nodes \textbf{SexNet} and then builds a network guided by the $BJD$ matrix.  The nodes with the smallest degree have the least flexibility, so we start building \se~  by randomly selecting a man node of \bso~ with the highest (social) degree and assign its desired sexual degree to be column size of $BJD$ matrix.  This is represented by \emph{stubs}, or unconnected edges, associated with this node. 

 We repeat the following process until all edges in \se~ -that are equal to the summation of elements in $BJD$ matrix- are placed: at any step, we find a node with the highest stub in \se~ and then with probability of $p$, we find a partner with proper degree defined by $BJD$  for them from their social contact in \bso~, or with the probability of $1-p$, we find a partner with proper degree defined by $BJD$  from closest people but not their social contact, and then reduce its new partner's stub by one. If we find all the partners for all nodes in \textbf{SexNet}, we introduce a new node from highly social active nodes in \bso~ to \se~ and assign its desired degree and stub equal to the current maximum degree frequency of \textbf{SexNet}.
 
To keep or remove an edge,  we have to calculate the degree of nodes attached to it for each possible edge in the \textbf{SocNet}, thus,  the full set of experiments run in O($|E|P^mP^w$) time, where $|E|$ is the number of edges in \so, $P^m$ number of its men nodes and $P^w$ number of its women nodes. This method is feasible if the average degree ($\frac{2|E|}{P^m+P^w}$) of the network is not high.

For completeness, we provide pseudo-code for our Python scripts in Algorithms \ref{main_alg} and \ref{help_alg}. 
Table  \ref{notation} is the table of symbols used in these Algorithms.
\begin{table}[htp]
\centering
\resizebox{\columnwidth}{!}
{
\begin{tabular}{ lp{15cm} }
\toprule[1.5pt]
   \textbf{Notation} & \textbf{Description}\\
  \cmidrule(lr){1-2}
 $G.n$ &set of nodes  in network $G$\\
 $G.e$ & set of edges  in network $G$\\
 $d_G$ & degree frequency list for network $G$\\
$d_G(\textbf{i})$ & degree of node \textbf{i} in network $G$\\
$G.N(\textbf{i})$ &  set of neighbors of node \textbf{i} in Network G\\
$dist(G,\textbf{u},\textbf{v})$ &  distance between two nodes \textbf{u} and \textbf{v} in Network G\\
$M.col~/~M.row$ &  column/row size of a matrix M\\
$M(i,:)~/~M(:,i)$ & $i^{th}$ row/column of matrix M\\
$V( i)$ &  $i^{th}$ element of vector V\\
$V.index(a)$ &  index of  element a in  vector V\\
$|S|$ &  size (the number of elements) of  a set S\\ 
$S.remove(m)$ & remove member m from   a set S\\ 
$S.sample(P)$ & randomly select an element with property P (if P$=1$ there is no property) from set S \\ 
$urn$ & uniform random number in $[0,1]$\\
\bottomrule[1.5pt]
\end{tabular}}
\caption{Table of notation for a conventional network $G$ in algorithms.}
\label{notation}
\end{table}

\resizebox{\columnwidth}{!}{\begin{minipage}{1.5\linewidth}
\begin{algorithm}[H]
\hspace{-.5cm} \textbf{function} \textbf{SexNet}=\sts(\bso,$BJD$, $p$)\\ 
\KwIn{%
         Revised social network \bso, bipartite joint-degree matrix $BJD$,
         $p \in[0,1]$ average  fraction of sexual partners selected from social contacts. 
         }%
\KwOut{Sexual network \textbf{SexNex}.}%
 \SetAlgoLined
{\textit{ \footnotesize{/* From the highest degree men nodes in \bso, randomly select one node to add to \textbf{SexNet}  */}}}\
 $\textbf{SexNet}.n=\emptyset,~\textbf{SexNet}.n \leftarrow \textbf{u}={max_{d_\so(\textbf{k})}}\{\textbf{k} \in \bso.n \}$  \;
 {\textit{ \footnotesize{/* The desired degree of the first selected node in \textbf{SexNet} is the column size of BJD */}}}\
 $d_{\se}(\textbf{u})\defeqq BJD.col,~stub(\textbf{u})\defeqq d_\se(\textbf{u})$ \;
 \textit{ \footnotesize{ /* $E$ is total number of edges in \textbf{SexNet} to be filled */}}\newline
 $E\defeqq \sum_i\sum_jBJD(i,j)$ \;
 \While{$|\textbf{SexNet}.e|\leq E$}{
 \textit{ \footnotesize{/* NF includes all the nodes in \textbf{SexNet} who are still looking for partners */}}\newline
 $NF\defeqq\{\textbf{k} \in \textbf{SexNet}.n~if~stub(\textbf{k})>0\}$\;
  \While {$|NF|\geq 1$}{
  \textit{ \footnotesize{/* Start with the highest degree nodes in $NF$ */}}\newline
  $\textbf{u}={max_{stub(\textbf{k})}}\{\textbf{k} \in NF \}$ \;
  {\textit{ \footnotesize{/* Find  partner for \textbf{u} via Algorithm \ref{help_alg} $\rightarrow$ partner \textbf{v} of degree $d'$ */}}}\newline
  $(d',\textbf{v})=\Call{FP}{\textbf{u}, \bso,            \textbf{SexNet}, BJD, NF, p}$\;
  \eIf{$(d',\textbf{v})= False$}{ 
 \textit{ \footnotesize{/* If could not find partner for \textbf{u} remove it from $NF$ */}}\newline
    $NF.remove$(\textbf{u}) \;
   }{ \textit{ \footnotesize{/* If find the partner \textbf{v}, add edge between \textbf{u} and \textbf{v} and reduce both their stubs by one */ }}\newline
   Add edge (\textbf{u},\textbf{v}) in \textbf{SexNet}$,stub(\textbf{u})\leftarrow stub(\textbf{u})-1,$ $stub(\textbf{v})\leftarrow stub(\textbf{v})-1$ \;
   \textit{ \footnotesize{/* If \textbf{u} [\textbf{v}] has found all the partners, remove it from the candidates in \textbf{SexNet} */}}\newline
    \If{stub(\textbf{u})=0 [stub(\textbf{v})=0]}
    {$d_\se.remove(d_\se(\textbf{u}))$ [$d_\se.remove(d_\se(\textbf{v}))$]\;}
    \textit{ \footnotesize{/* Update the corresponding entry in $BJD$ */}}\newline
    \eIf{\textbf{u} is woman}{$BJD(d,d')\leftarrow BJD(d,d')-1$}{$BJD(d',d)\leftarrow BJD(d',d)-1$}
  }}
   \textit{ \footnotesize{/* From the nodes in \bso~ but not \textbf{SexNet}, choose one node with highest social degree, add to \textbf{SexNet} */}}\newline
   \textbf{SexNet}$.n \leftarrow \textbf{u}={max_{d_\so(k)}}\{k \in \bso.n-\textbf{SexNet}.n\}$\;
    \textit{ \footnotesize{/* Define its degree and stub  equal to maximum value in degree frequencies of \textbf{SexNet} */}}\newline
   $d_\se(\textbf{u})\defeqq max\{d_\se\},~stub(\textbf{u})\defeqq d_\se(\textbf{u}).$
  }
  \textbf{return} \textbf{SexNet}\;
   \caption{ Extracting sexual network, \textbf{SexNet}, from a given social network, \textbf{BSocNet}}\label{main_alg}
\end{algorithm}
\end{minipage}}
\resizebox{\columnwidth}{!}{\begin{minipage}{1.5\linewidth}
\begin{algorithm}[H]
\hspace{-.5cm} \textbf{function} $(d',\textbf{v})$=FP(\textbf{u}, \textbf{SocNet}, \textbf{SexNet}, $BJD$, $NF$, $p$)\\ 
\KwIn{Node \textbf{u} in the sexual network \textbf{SexNet}, social network \textbf{SocNet}, $BJD$ matrix, set $NF$ of nodes who need partners, $p\in[0,1]$ fraction of seuxal partnerships from social network.
        }%
\KwOut{($d'$,\textbf{v}):  Node \textbf{v} with its desired  degree $d'$.}%
 \SetAlgoLined
   $d=d_{\se}(\textbf{u})$\;
    \textit{ \footnotesize{/* Depending on sex of node and its degree, take the $d^{th}$ row or column of $BJD$ */}}\newline
      \lIf{\textbf{u} is woman}
       {$R\defeqq BJD(d,:)$, \textbf{else} $R\defeqq BJD(:,d)$}
\For{range($|R|$)}
{\eIf{$R\neq0$}
    {\textit{ \footnotesize{/* Choose one nonzero element of $R$ with index $d'$, which is the degree of partner to be found */}}\newline
    $w=R.sample(R(d')\neq 0)$, $d'=R.index(w)$\;
    \eIf{$urn\leq$p}
          {\textit{ \footnotesize{/* Prepare for choosing from the social contacts: */ \\/* $K1$ - set of available nodes in \textbf{SexNet} with degree $d'$, who are also social contact; */\\[-0.5em] /* $K2$ - set of nodes that are not in \textbf{SexNet} but the social contacts of \textbf{u} with  social degree $\geq d'$ */}}\newline
          $K1\defeqq\{\textbf{k}\in NF: \textbf{k}\in \textbf{SocNet}.N(\textbf{u})-\textbf{SexNet}.N(\textbf{u}), d_\se (k)=d'\}$\;\vspace*{-.0cm}
         ~$K2\defeqq\{\textbf{k}\in \textbf{SocNet}.N(\textbf{u})-\textbf{SexNet}.n:  d_\so(k)\geq d'\}$\;} 
         {\textit{ \footnotesize{/* Prepare for choosing outside the social contacts: */\\ 
         /* $K1$ - set of available nodes in \textbf{SexNet} with sexual degree $d'$, who are not social contacts */\\[-0.5em]
         /* $K2$ - set of nodes that are not \textbf{SexNet} and not social contact of \textbf{u} with social degree $\geq d'$)*/}}\newline 
         $K1\defeqq \{\textbf{k}\in NF: \textbf{k}\notin \textbf{SocNet}.N(\textbf{u})\cup\textbf{SexNet}.N(\textbf{u}), d_\se (k)=d'\}$\;
        ~$K2\defeqq \{\textbf{k}\in \textbf{SocNet}.n-\textbf{SexNet}.n:  d_\so(k)\geq d'\}$\;}
         
          \uIf{$K1\neq \emptyset$}
           {\textit{\footnotesize{/* If $K1$ is not empty, select the closest node in \textbf{SocNet} */}}\newline
           \footnotesize{\textbf{v}$=K1.sample(\textbf{w}: dist(\textbf{SocNet},\textbf{u},\textbf{w})=min\{dist(\textbf{SocNet},\textbf{u},\textbf{k})~for~k\in K1\})$}\;
              Break\; }
              \uElseIf{$K2\neq \emptyset$}
              {\textit{\footnotesize{/* if $K2$ is not empty, select the closest node in \textbf{SocNet}, define its sex degree and stub as $d'$ */}}\newline
              \footnotesize{\textbf{v}$=K2.sample(\textbf{w}: dis(\textbf{SocNet},\textbf{u},\textbf{w})=min\{dis(\textbf{SocNet},\textbf{u},\textbf{k})~for~k\in K2\})$\;
       $d_\se(\textbf{v})\defeqq d'$, $stub(\textbf{v})\defeqq d'$}\;
          Break\;}
              \Else{\textit{ \footnotesize{/*  If both don't work, move on to a different degree */}}\newline$R(d')\defeqq 0$\;}
         }
         {\textit{ \footnotesize{/* If all elements of $R$ are 0, we fail to find a proper partner for \textbf{u} */}}\newline$(d',\textbf{v})=False$\;
         break\; }
         }
\textbf{Return} $(d',\textbf{v})$\;
 
   \caption{  Finding partner with proper degree for a given node (\textbf{FP})}\label{help_alg}
\end{algorithm}
\end{minipage}}
\newpage


The initial algorithm will generate a network with the prescribed degree distribution.  We have observed that the resulting network always had the desired $BJD$ matrix for all of the sparse heterosexual networks we have generated in this project.  
However, there are some situations, where the desired network is not sparse, the algorithm can fail to exactly produce a network with the desired $BJD$ matrix for the joint-degree distribution.
When this happens, a second algorithm is used to rewire the network so that it will exactly match the desired bipartite joint-degree distribution for the number of partners that a person's partners have. 
\subsubsection{Rewiring \se~ for a given {$BJD$} matrix}
The rewiring  algorithm corrects any mismatch between the joint-degree distribution of generated \se~ and the desired $BJD$.  We define the joint-degree distribution of the generated \se~ as $\widetilde{BJD}$ and the mismatch error matrix $\mathcal{E}=BJD-\widetilde{BJD}$. If the matrix  $\mathcal{E}$ has nonzero elements, then the network is rewired to eliminate the error.  There are three possible cases:
\begin{enumerate}[(i)]
\item If entry $\mathcal{E}_{(i,j)}, (i, j>1)$, is a positive value $k$, it means that \se~ needs $k$ more edges between degree $i$ women and degree $j$ men. To create these edges,  we iterate the following process $k$ times: 
\begin{enumerate}[(a)]
\item First, identify a woman, \textbf{w}$_i$,  in \se, where $d_{\textbf{SexNet}}(\textbf{w}_i)=i$, and has as a partner, \textbf{m}$_1$,  with degree-1, i.e. $d_{\textbf{SexNet}}(\textbf{m}_1)=1$.  
\item Next, identify another man, \textbf{m}$_j \in \textbf{SocNet}.N(\textbf{w}_i)-\textbf{SexNet}.N(\textbf{w}_i)$,  
where  $d_{\textbf{SexNet}}(\textbf{m}_j)=j$  and \textbf{m}$_j$  has a degree-1 partner, \textbf{w}$_1$.  That is $d_{\textbf{SexNet}}(\textbf{w}_1)=1$. 
\item Finally, we rewire the network by removing the edges (\textbf{w}$_i$,\textbf{m}$_1$) and (\textbf{m}$_j$,\textbf{w}$_i$) and add edge (\textbf{w}$_i$, \textbf{m}$_j$), as illustrated in the Rewiring (1) of the Figure (\ref{rewiring}).
\end{enumerate}

\item If element $\mathcal{E}_{(i,j)}$, $(i, j>1)$, is a negative value $k'$, it means that \se~ have extra $k'$ edges between degree $i$ women and degree $j$ men. To remove these edges, we iterate following  process $k'$ times: 
\begin{enumerate}[(a)]
\item First, identify a woman, \textbf{w}$_i$,  in \se, where $d_{\textbf{SexNet}}(\textbf{w}_i)=i$, which has a degree-j partner like \textbf{m}$_j$, that is $d_{\textbf{SexNet}}(\textbf{m}_j)=j$. The nodes \textbf{w}$_i$ and \textbf{m}$_j$ are selected so that they have at least one social contact  with opposite sex that is not their sexual partner. 
\item Next, identify another man, \textbf{m}$_1\in \textbf{SocNet}.N(\textbf{w}_i)-\textbf{SexNet}.N(\textbf{w}_i)$, and a woman, \textbf{w}$_1\in \textbf{SocNet}.N(\textbf{m}_j)-\textbf{SexNet}.N(\textbf{m}_j)$. 
\item Finally, rewire the network by removing the edge (\textbf{w}$_i$,\textbf{m}$_j$) and add edges (\textbf{w}$_i$, \textbf{m}$_1$) and (\textbf{w}$_1$,\textbf{m}$_j$), as illustrated in the Rewiring (2) of the Figure (\ref{rewiring}).
\end{enumerate}

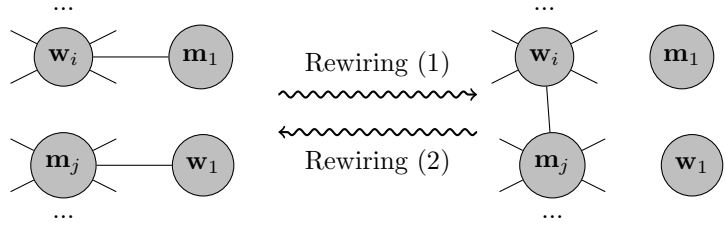
\begin{figure}[htp]
\centering
\begin{tikzpicture}
    \node[draw=black, circle, fill=gray!50] (wi) at (0,0) {$\textbf{w}_i$};
    \draw let \p1=(wi) in (wi) -- (\x1+20,\y1+10);
    \draw let \p1=(wi) in (wi) -- (\x1-20,\y1+10);
    \draw let \p1=(wi) in (wi) -- (\x1+20,\y1-10);
     \draw let \p1=(wi) in (wi) -- (\x1-20,\y1-10);
     \node[above= .1cm of wi]{...};
\node[draw=black, circle, fill=gray!50, right=1cm of wi] (m1) {\textcolor{black}{\textbf{m}$_1$}};
     \draw [] (wi) to node {} (m1);   
     \node[draw=black, circle, fill=gray!50, below=1cm of wi] (mj) at (0,0) {$\textbf{m}_j$};
    \draw let \p1=(mj) in (mj) -- (\x1+20,\y1+10);
    \draw let \p1=(mj) in (mj) -- (\x1-20,\y1+10);
    \draw let \p1=(mj) in (mj) -- (\x1+20,\y1-10);
     \draw let \p1=(mj) in (mj) -- (\x1-20,\y1-10);
     \node[below= .1cm of mj]{...};
\node[draw=black, circle, fill=gray!50, right=1cm of mj] (w1) {\textcolor{black}{\textbf{w}$_1$}};
     \draw [] (mj) to node {} (w1); 
     
       \node[draw=black, circle, fill=gray!50, right=6cm of m1] (rwi) at (0,0) {$\textbf{w}_i$};
    \draw let \p1=(rwi) in (rwi) -- (\x1+20,\y1+10);
    \draw let \p1=(rwi) in (rwi) -- (\x1-20,\y1+10);
    \draw let \p1=(rwi) in (rwi) -- (\x1+20,\y1-10);
     \draw let \p1=(rwi) in (rwi) -- (\x1-20,\y1-10);
     \node[above= .1cm of rwi]{...};
\node[draw=black, circle, fill=gray!50, right=1cm of rwi] (rm1) {\textcolor{black}{\textbf{m}$_1$}};
    \node[draw=black, circle, fill=gray!50, below=1cm of rwi] (rmj) at (6.5,0) {$\textbf{m}_j$};
    \draw let \p1=(rmj) in (rmj) -- (\x1+20,\y1+10);
    \draw let \p1=(rmj) in (rmj) -- (\x1-20,\y1+10);
    \draw let \p1=(rmj) in (rmj) -- (\x1+20,\y1-10);
     \draw let \p1=(rmj) in (rmj) -- (\x1-20,\y1-10);
     \node[below= .1cm of rmj]{...};
\node[draw=black, circle, fill=gray!50, right=1cm of rmj] (rw1) {\textcolor{black}{\textbf{w}$_1$}};
     \draw [] (rmj) to node {} (rwi);
  \draw [-to,thick,snake=snake,segment amplitude=.4mm,segment length=2mm,line after snake=1mm]
    ([xshift=10mm,yshift=-5mm]m1 -| w1) -- ([xshift=-10mm,yshift=-5mm]rwi -| rmj)
    node [above=1mm,midway,text width=3cm,text centered]
      { Rewiring (1) };
      \draw [-to,thick,snake=snake,segment amplitude=.4mm,segment length=2mm,line after snake=1mm]
      ([xshift=-10mm,yshift=-10mm]rwi -| rmj) -- ([xshift=10mm,yshift=-10mm]m1 -| w1)
    node [below=1mm,midway,text width=3cm,text centered]
      { Rewiring (2) };
\end{tikzpicture}
\vspace*{8pt}
\caption{Schematic of steps 1 and 2 of Rewiring approach to correct $BJD$.}
\label{rewiring}
\end{figure}
\item In the previous steps, we pushed  back nonzero elements in $\mathcal{E}$ to its first row and column, which causes new nonzero elements in the first row and column.
To remove these nonzero values, we have to add or remove small components. For example, if the element (i,1) of $\mathcal{E}$ is a positive value $k$, it means  that we need a small component of a degree $i$ woman whose partners  are all degree 1 men. Therefore, we simply make this component from the people who are not currently in \textbf{SexNet}$.n$. If the element $(1,j)$ of $\mathcal{E}$ is a negative value $k$, it means  we have to remove a small component of a degree $j$ man whose partners are all degree 1 women. Therefore, we simply look for such a component and remove it from \textbf{SexNet}.
\end{enumerate}

We have found that  this algorithm almost always converges to the desired $BJD$.  However, there are rare cases when the desired rewiring nodes may not exist, and algorithm stalls with  $\mathcal{E} \ne 0$.  When this happens, the rewired network will have a better joint-degree distribution of \textbf{SexNet}, even though it does not exactly match the desired $BJD$.  In the numerical simulations, all of the generated New Orleans heterosexual network had exactly the desired $BJD$ based on survey data. 

In the next section, we apply our algorithm to generate and analyze several random \se~ corresponding to sexual activity of adolescent and young adult sexually active African Americans reside in New Orleans. First, we explain the inputs of our approach: New Orleans social network and joint-degree distribution- $BJD$ matrix- corresponding to its sexual network. Then we use our algorithm to generate and analyze a bunch of \se s for a subpopulation of people in New Orleans.

\section{Simulations}

We analyze an ensemble of sexual networks with a prescribed joint-degree distribution representing sexual activity of young adult African Americans in New Orleans.

\subsection{The New Orleans social  activity data and \so} 
The \so~ is based on the synthetic data generated by Simfrastructure \cite{eubank2010detail, eubank2008synthetic}
for 130,000 synthetic people residing in New Orleans.
Simfrastructure is a high-performance, service-oriented, agent-based simulation system, representing and analyzing interdependent infrastructures.  
The data for the network includes information for each individual, identified by their \textbf{PID} (personal identifier),  and includes their age,  gender, household, and other demographic information.  The contact information for  each PID is encoded in the contact file in Table \ref{cfile}. 
 
\begin{table}[h]
\centering 
\scalebox{.8}{\begin{tabular}{c rrrr} 
\toprule[1.5pt]
\textbf{PID} & \textbf{FID}& \textbf{A}& \textbf{T} \\ 
\hline
43722 & 16981& H &$0.3$ \\ 
 & 11462& W& $0.2$ \\ 
 & 37790& Sh& $0.01$ \\ 
  & \vdots & \vdots & \vdots \\
  \hline
51981 & 23476& Sc&$0.16$ \\ 
 & 18462& O& $0.02$ \\ 
 & 10790& H& $0.5$ \\ 
  & \vdots & \vdots & \vdots \\ \hline
 \vdots & \vdots & \vdots & \vdots \\ 
\bottomrule[2pt]
\end{tabular}}
\caption[\textbf{The contact file of social network}]{Table input for the social contact network of individuals. The \textbf{PID} is personal ID, \textbf{FID} is their social contact ID, \textbf{A} is activity in which PID meet FID, and \textbf{T} is the fraction of  time in a day that two social contacts meet with each other through activity A. We have five different activities, including H as home, W as work, Sc as school, Sh as shopping, and O as others. For example, person 43722 stays with person 16981 at the same home for $7.2$ hours in a day.
}
\label{cfile}
\end{table}

The Simfrastructure data was used to generate the original \so, which was then used to generate \eso~ and \bso~ as described in the previous section.

\subsection{The {$BJD$} matrix for New Orleans heterosexual activity }
An ongoing community-based pilot study was conducted among sexually active African Americans ages $15-25$ in New Orleans \cite{kissinger2014check}, to assess the effectiveness of prevention and intervention programs for  chlamydia.
Socio-demographic information including age, race, educational level and, sexual behavior--number and age of heterosexual partners  in the past two months-- and history of their STI test results were collected from $202$ men and $414$ women participants. Meanwhile, their partners' information  has been collected by asking questions referring to the status of each relationship such as the partner's age  and the possibility that their partner(s) have intercourse with others.  The survey results were used to 
construct the $BJD$ matrix of a heterosexual network of individuals in New Orleans. For a population $P=15,000$ sexually active young adult men and their women partners residing in New Orleans, we have

\vspace{.5cm}
\resizebox{\columnwidth}{!}{$BJD_{15000} = 
\left( 
{\begin{array}{*{21}c}
 1663 & 1588 &  1225 & 896  & 645 &469& 342  &252 &186 &138 &105 &90&72&57&41 &34  &23 &24&14&13&14\\
474  &  452 & 350   & 255  & 185 & 133&97   &73  &52  &39  &31   &26&20&17&11&9&8&7&3&4&4\\
198  &  188 & 145   & 107  & 77  &57 &40   &29  &22  &16  &12   &12&9&6&5&3&2&4&1&1&2\\
68   &  63  & 49    & 35   & 26  & 18&15   &10  &8   &6   &4    &3&3&3&1&2&0&0&1&10&0\\
18   & 18   & 13    &9     & 6   &6 &3    &4   &1   &1   &2    & 0&0&1&1&0&1&0&0&1&0\\
3    & 3    &  3    & 2    & 1   &1 &0    &0   &1   &0   &0    & 1&0&0&1&0&0&1&0&0&1\\   
 \end{array}  }\right).
$}

\vspace{1cm}
The dimension of this $BJD$ matrix is $6\times21$, that is, the maximum number of partners women have is $6$ and for men is $21$. 

\subsection{\se~ analysis}\label{NA}
Using the social network and $BJD$ matrix provided in the previous subsections and approach described in Section \ref{method}, we generated $150$  \se s  of $15000$ people for $p=0.2,0.4,0.6,0.8$ and $1$. That is, $30$ of the \se s are $20\%$, $30$ are $40\%$, $30$ are $60\%$, $30$ are $80\%$, and the rest $30$ are $100\%$ subgraph of \bso. We then compared some descriptive measures of  this ensemble of random networks  that were not imposed when generating the networks, including the size of giant components and bi-components, number of connected components, and average redundancy coefficient.

\begin{figure}[htp]
\begin{center}
\includegraphics[width=.85\textwidth]
{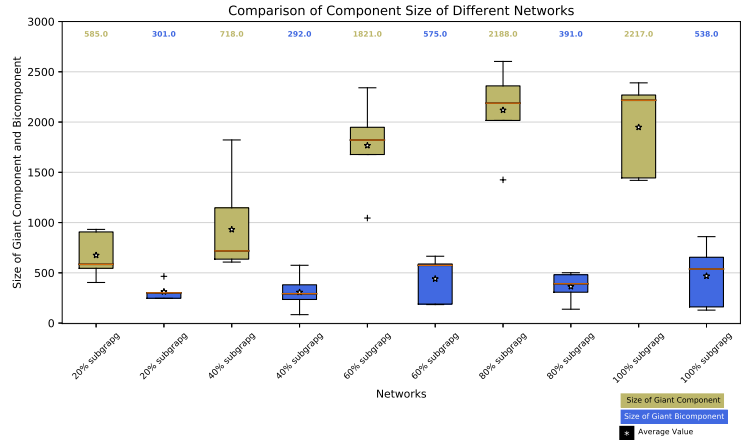}
\vspace*{8pt}
\caption{Box plot representing thee size of the giant component and bi-component for each group of networks: The size of the giant component becomes bigger when the portion of the subgraph becomes stronger, however, the social network does not have much impact on the size of the giant bi-component. }
\label{fig:sg}
\end{center}
\end{figure}

First, we evaluated and compared the size of the giant component and bi-component (the first and second biggest connected components of the network) for each group of the networks. Figure (\ref{fig:sg}) shows the box plot of these sizes: there is an increment in the size of giant components when people select most of their  sexual partners from their social contacts. Because in that case, sexually active people are tighter together within the social contact network. But there is not a significant difference in the size of giant bi-component.

\begin{figure}[htp]
\begin{center}
\includegraphics[width=.85\textwidth]{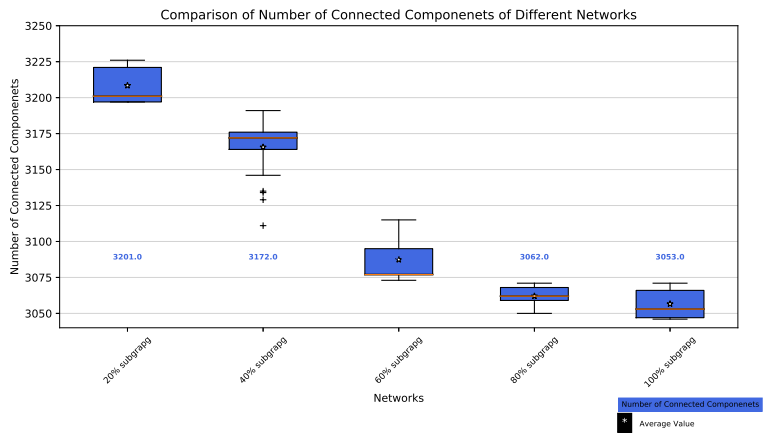}
\caption{Box plot representing the number of  connected components, $Nc$, for each group of networks: A significant difference is observed  in $Nc$ between each group, $Nc$ is lower in larger subgraph of social networks. }
\label{fig:nc}
\end{center}
\end{figure}

The number of connected components, $Nc$, is another measure characterizing  network toughness. This measure can be, not necessarily, correlated to  the component's size of the network.  Figure (\ref{fig:nc})  displays descriptive statistics for $Nc$ in each network group. Note that  data distributions are approximately symmetrical, and measures of $Nc$ are similar across groups, but, they change  by changing the source of partner selection- changing $p$.

Redundancy coefficients are the measure of the degree to which nodes in a bipartite graph tend to cluster together:

\begin{definition} For a bipartite network,  redundancy for a node  is the ratio  of its overlap  to its maximum possible overlap  according to its degree.
The overlap of a node is the number of pairs of neighbors that have mutual neighbors themselves, other than that node \cite{latapy2008basic}. For a typical node \textbf{v}, the redundancy coefficient of \textbf{v} is defined as 

$$Rc(\textbf{v})=\frac{|\{\{\textbf{u,\textbf{w}}\}\subseteq N(\textbf{v}), \exists \textbf{v'}\neq \textbf{v}~s.t~\textbf{uv'}\in\textbf{E}, \textbf{wv'}\in \textbf{E} \}|}{\frac{|N(\textbf{v})|(|N(\textbf{v})|-1)}{2}},$$

where, $N(\textbf{v})$ is the set of all neighbors of node \textbf{v}, and \textbf{E} is the set of all edges in the network.
\end{definition}

\begin{figure}[htp]
\begin{center}
\includegraphics[width=.85\textwidth]{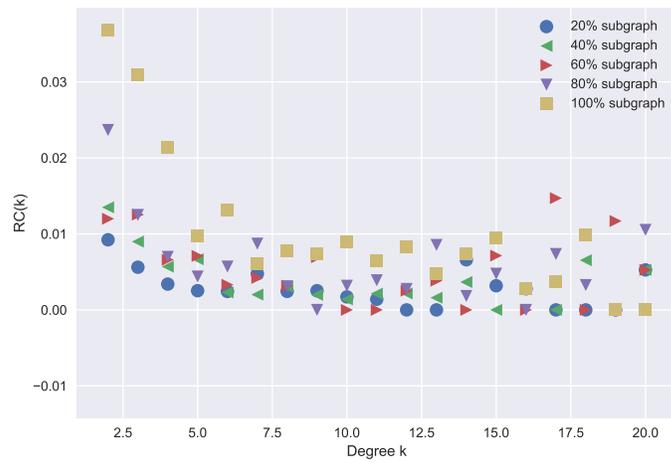}
\caption{Scatter plot of $Rc$  versus degree for five different networks:  $Rc$ for \se~  which is  strong subgraph of \bso~ (higher $p$) is higher, because clustering coefficient for \bso~ is high and therefore, \se~ inherits this property by having higher $Rc$ than the ones which are weak subgraph of \bso. }
\label{cl_rc}
\end{center}
\end{figure}

We compare this measure for the networks in Figure (\ref{cl_rc}):  each data point $Rc(k)$ for degree $k$ is obtained by averaging  redundancy coefficient over the group of  people with $k$ partners. In most of the networks, $Rc(k)$ decreases with $k$ \cite{newman2010networks}.  Redundancy coefficient $Rc$ is affected by social network \bso: when people select more sexual partners from their social contacts the value for $Rc$ increases, which  is because of stage one  of the algorithm- Generate an extended social network. In that stage, by  connecting the social contacts of a person in \bso, we increase its clustering coefficient. Therefore, because increasing $p$ \se~ becomes a stronger subgraph of \bso, it inherits more properties from \bso, that is, by increasing $p$  $Rc$ of the \se, which is correlated to clustering coefficient of \bso, increases.

\section{Discussion} 
We described a new algorithm to generate an ensemble of heterosexual networks based on heterosexual behavior surveys  for the young adult African American population in New Orleans.   The prescribed degree and joint-degree distribution represented the heterosexual network embedded within a social network that captures the biased mixing of the population based on age,  physical location, and social activities. 

We generated an ensemble of different heterosexual networks with the same $BJD$.
When the networks had a more percentage of partners selected from their extended social contacts, then we 
observed a tighter distribution in the number of connected components and the size of giant component and bi-components.
When more partners are chosen from the extended social network, instead of randomly selected from the population, then the size of giant component increases,  
 and following that the number of connected components decreases, which  is because of reducing the mixing in generating sexual network: when people select their sexual partners from their social contacts they stand in a tight group within social network. In fact, when being subgraph of the extended social network become stronger, the candidate set of sexual partners for each person that is  set of social contacts decreases and becomes local ( this set includes close contacts and contacts of contacts) compared with when this set is the whole population.


As $p$ increases, then more partners are chosen from a person's extended social network.  This also increases the network clustering coefficients (where more partners of your partner's partner are also one of  your partners). 
The redundancy coefficients for networks increases as the dependence of sexual  network  on social one rises when $p$ increases, which is because of the high clustering coefficient of the social network due to the first stage of the algorithm, generate an extended social contact network. In that stage, we made some new contacts between the contacts of each individual, which causes the increment in the clustering coefficient of the social contact network.
On the other hand, when more partners are chosen from the social contact network, more properties of the social network such as the clustering coefficient become inherited by \se. Thus, increasing $p$, we observe increment in the redundancy coefficient of \se.

We studied the measures of networks because they may affect the spread of an STI such as chlamydia on the \se s. In our future work when studying the spread of chlamydia on heterosexual networks, we will measure their impact on  the prevalence of chlamydia over \se s generated using different $p$ values.

There are still unanswered questions for proving the existence of a heterosexual network with a prescribed joint-degree distribution embedded within a prescribed social network.   That is, there are no explicit criteria to guarantee that a  heterosexual network with a particular joint-degree distribution can be embedded within a particular social network or not.  
 
We are currently simulating a stochastic agent-based network model on \se~ for the spread of chlamydia and comparing different intervention strategies to control the spread of STIs.  These simulations will use the underlying social contact network to improve the current intervention  models by considering  the impact of counseling and behavioral changes such as increasing condom use or  social contact notification.

\section*{Acknowledgments}
The authors thank Achla Marathe, Patricia Kissinger, Stephen Eubank, and Norine Schmidt  for their useful comments and suggestions. 
This work was supported by the endowment for the Evelyn and John G. Phillips Distinguished Chair in Mathematics at Tulane University and grants from the National Institutes of Health National Institute of Child Health and Human Development (R01HD086794) and Office of Adolescent Health (TP2AH000013)   and the 
National Institute of General Medical Sciences program for Models of Infectious Disease Agent Study (U01GM097658).
The content is solely the responsibility of the authors and does not necessarily represent the official views of the National Institutes of Health.

\end{document}